\documentclass[epj]{svjour}

\usepackage{graphicx}
\usepackage{rotating}
\usepackage{dcolumn}

\begin{document}
\title{Nanorheology : an Investigation of the Boundary Condition at Hydrophobic and Hydrophilic Interfaces.}

\author{C. Cottin-Bizonne \inst{1} \and S. Jurine \inst{1} \and J. Baudry
\inst{1}
 \and J. Crassous \inst{2}  \and F. Restagno \inst{1,} \inst{3} \and \'E. Charlaix
\inst{1}
}                     
\offprints{ccottin@dpm.univ-lyon1.fr}          

\institute{D\'epartement de Physique des Mat\'eriaux (UMR 5586),
Universit\'e Lyon I, 43 bd du 11 Novembre 1918, 69622 Villeurbanne
Cedex (France)  \and
 Laboratoire de Physique (UMR 5672), ENS-Lyon, 46
all\'ee d'Italie, 69364 Lyon Cedex 07 (France) \and Now at
Laboratoire des Fluides Organisé\'es, Coll\`ege de France, 11,
place M. Berthelot, 75231 Paris cedex 05 (France)}
%
%
\abstract
    {It has been shown that the flow of a simple liquid over
a solid surface can violate the so-called no-slip boundary
condition. We investigate  the flow of polar liquids, water and
glycerol, on a hydrophilic Pyrex surface and a hydrophobic surface
made of a Self-Assembled Monolayer of OTS
(octadecyltrichlorosilane) on Pyrex. We use a Dynamic Surface
Force Apparatus (DSFA) which allows one to study the flow of a
liquid film confined between two surfaces with a nanometer
resolution. No-slip boundary conditions are found for both fluids
on hydrophilic surfaces only. Significant slip is found on the
hydrophobic surfaces, with a typical length of one hundred
nanometers. 
}

\maketitle

\section{Introduction}
\label{intro}

The hydrodynamic properties of a liquid at a solid interface are
very important in a number of applications involving flows in
confined geometries: tribology and lubrication, flows through
porous media, or in microfluidic devices, \textit{etc}. At a
macroscopic scale, the no-slip boundary condition widely used to
describe the flow of simple liquids at a solid surface is usually
considered as very robust. However, it was suggested in early
works of Schnell \cite{schnell56} and later of Churaev \textit{et
al.} \cite{Churaev84} that simple liquids may undergo substantial
slip when flowing on non wetting surfaces. By measuring the water
pressure drop in silanized glass capillaries of micrometric size
the latter showed slip effects with a typical length of tens of
nanometers. Such effects can not be neglected in flows at small
scale, and the recent development of microfluidic devices has led
to an increasing interest on this subject
\cite{tretheway2002,cheng2002}.

At a microscopic scale, the hydrodynamic boundary condition (HBC)
at a solid surface reflects the transfer of tangential momentum
between the two phases. The influence of the liquid-solid
interactions on the HBC has been addressed theoretically. These
interactions can be characterized by the contact angle of the
liquid on the solid surface.  Molecular dynamic simulations by
Barrat and Bocquet \cite{Barrat99a,Barrat99b} of Lennard-Jones
liquids have shown that, when the liquid wets the solid surface,
the no-slip boundary condition holds. On the contrary, in the case
of partial wetting, slip effects of several molecular diameters
occur: for a contact
 angle of $140^{o}$, slipping lengths of about 10~nm are found
\cite{Barrat99a,Barrat99b}. At high shear rate, ($\dot \gamma \geq
0.2\tau$, where $\tau$ is the characteristic time of the
Lennard-Jones potential) it has also been shown that the HBC may
become non-linear and rate dependent \cite{Thompson97}. From an
experimental point of view, investigations of flows at a
submicronic scale have developped with experimental techniques
such as Surface Force Apparatus (SFA), modified Atomic Force
Microscopy (AFM) and fluorescence recovery methods. For a simple
Newtonian liquid, it has
 been shown that the flow in films thicker than about ten
molecular diameters can be described by the Navier-Stokes
equations \cite{Horn85,Georges93,Klein2002}. The first experiments
with SFA have also shown that, in a number of cases where the
liquids wet the solid surfaces, the no-slip boundary condition
holds. Studying the flow of alkanes over metallized surfaces of
roughness close to 1~nm, Georges \textit{et al.} \cite{Georges93}
have shown that the no-slip boundary condition should be applied
at, or very close to, the solid surface, at the scale of a
molecular diameter. Previously, Chan and Horn \cite{Horn85} have
obtained the same results for confined non polar liquid between
atomically smooth mica surfaces. More recently, different research
groups have studied various liquid-solid systems in both wetting
and non-wetting conditions. They have reported various results
either in the nature of the HBC or in the magnitude of the effects
observed. We will summarize in Part~V of this paper a number of
such recent results.

In this paper, we use a Dynamic Surface Force Apparatus (DSFA) to
compare the flow and the HBC of glycerol and water confined
between hydrophilic (Pyrex) surfaces and hydrophobic silanized
Pyrex surfaces.  This SFA creates an
 oscillatory flow of very small amplitude and variable frequencies
  of a liquid confined between a sphere and a plane.
The HBC is deduced from the force measured on the plane, caused by
the flow of the liquid squeezed between the surfaces.
  We will first describe the dynamic SFA, then the experimental system
 and finally present our results. These are compared to previous
 measurements by other groups with different experimental
 techniques.

\section{The experimental setup}
\label{sec:expsetup}

\begin{figure}[htbp]
\begin{center}
\includegraphics[width=8cm]{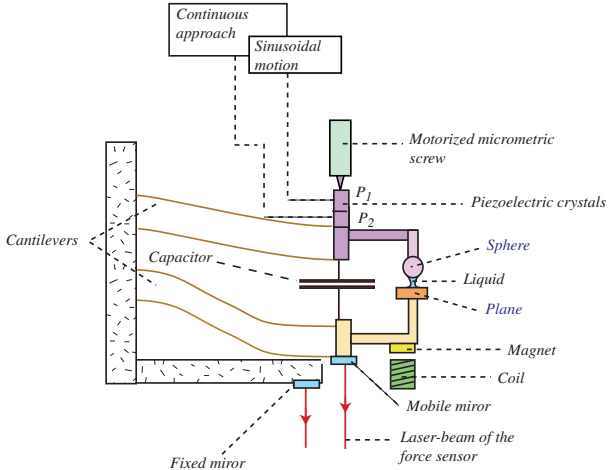}
\caption{Schematic illustration of the Dynamic Surface Force
Apparatus} \label{schemprinc}
\end{center}
\end{figure}

The experimental setup is a new Dynamic Surface Force Apparatus
(DSFA) that has recently been developed by our group.  A schematic
representation is given in Fig.\ref{schemprinc} and a more
detailed description can be found in previous articles
\cite{restagno2001,rsi2002}. This DSFA measures separately the
relative displacement $h$ of the surfaces and the interaction
force between them. The surfaces used are a plane and a sphere.
The apparatus measures the static as well as the dynamic component
of the forces when the sphere is moved towards the plane and  can
therefore be used as a nanorheometer. The sphere can be moved in a
direction normal to the plane over a distance of about 2~$\mu$m
with a piezoelectric element $P_{1}$. A piezoelectric element
$P_{2}$ is used to superimpose to the quasistatic
 motion of the sphere a sinusoidal displacement of small amplitude at a
frequency $\omega/2\pi$ between 5 and 200 Hz.

In the  experiments reported here, a planar surface is fixed on a
double spring cantilever of stiffness 1700~N/m. Its displacement
is measured with a Nomarski interferometer, with a static
resolution of 0.5~\AA~corresponding to 80~nN. The quasistatic
displacement between the sphere and the plane is measured with a
capacitive sensor with a 1~\AA~resolution. This capacitive sensor
is calibrated for static and dynamic displacements before each
experiment, with the help of the interferometer and of a
coil/magnet system. The relative displacement of the surfaces is
thus directly measured without using the calibration of the
piezoelectric drivers. For dynamic measurements, the output
signals of the sensors are connected to two digital two-phase
lock-in amplifiers (Standford Research System SR830 DSP Lock In
Amplifier) whose reference is used to drive the piezoelectric
$P_{2}$. In the dynamic regime, the capacitive sensor has a
sensitivity of 2.10$^{-12}$~m.Hz$^{-1/2}$ and the optic sensor has
a sensitivity of 8.10$^{-9}$~N.Hz$^{-1/2}$. Our measurements are
performed with a bandwidth of 1 Hz. The phase resolution is
0.5$^{{o}}$. The dynamic transfer fonction of the apparatus itself
is measured for all frequencies when the surfaces are very far
apart (no force). The dynamic component of the force is then
deduced from the experimental signal, using this transfer fonction
\cite{rsi2002}.

In this paper we will use the following notations: the
displacement between the surfaces is
\begin{equation}
h(t)=h_{{dc}}(t)+Re[\widetilde{h}_{{ac}}e^{\textrm{\small{j}}\omega
t}]
\end{equation}
where $h_{{dc}}$ is the quasistatic displacement of the surfaces
with constant velocity (usually 1~\AA/s) and
$\widetilde{h}_{{ac}}$ the complex amplitude of the oscillatory
component of the displacement. The interaction force is:
\begin{equation}
F(t)=F_{{dc}}(t)+Re[\widetilde{F}_{{ac}}e^{\textrm{\small{j}}\omega
t}]
\end{equation}
where $F_{{dc}}$ is the quasistatic force of the surfaces and
$\widetilde{F}_{{ac}}$ the complex amplitude of the oscillatory
component of the force.

 The norm and phase of the complex amplitudes
$\widetilde{h}_{{ac}}$ and $\widetilde{F}_{{ac}}$ are determined
using the procedure described above. The transfert function is
defined as follows:
\begin{equation}
\widetilde{G}(\omega)\equiv\frac{\widetilde{F}_{{ac}}}{\widetilde{h}_{{ac}}}=
G^{\prime}(\omega)+\textrm{j}G^{\prime\prime}(\omega)
\end{equation}
where $G^\prime(\omega)$ and  $G^{\prime\prime}(\omega)$ are
respectively the stiffness and the damping induced by the confined
liquid.

The signal $h_{{dc}}$ being a displacement, it is necessary to
specify the origin in order to get the actual distance between the
surfaces. This origin is usually determined using the theory of
adhesive contact (the JKR theory \cite{Johnson71}). In this paper,
the origin is only determined as the point where the repulsive
force between the surfaces starts to raise (this only makes a
difference of about 2 nm with the JKR determination, and
sufficiently precise for our experiments ).

\section{The experimental system} \label{sec:syst}

We have studied three systems with different wetting properties:
glycerol on Pyrex, glycerol
 on silanized Pyrex, and water on silanized Pyrex.

\begin{figure}
\begin{center}
\includegraphics*[height=6cm]{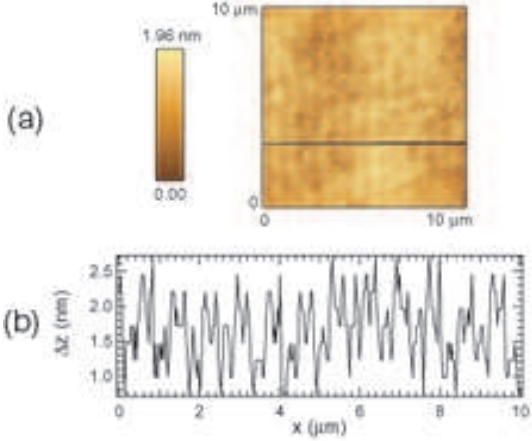}
\caption{AFM image of a Pyrex plane} \label{AFM}
\end{center}
\end{figure}

 The surfaces, a sphere with a millimeter diameter and a plane, are prepared as follows.
The surfaces are washed in an
 ultrasonic bath with a detergent and distilled water. Then,
 they are rinsed with purified propanol, and, finally, they are passed
 through a flame. We have studied the surface of the plane with an
atomic
 force microscope (Fig.\ref{AFM}). The peak to
 peak roughness of the surfaces prepared in this way is about 1~nm on
a surface of $10\mu$m$\times 10\mu$m. The sphere is always plain
Pyrex.
The silanized planes have been prepared by submerging
 our Pyrex surfaces (washed as explained above) in a
 mixture of 100~$\mu$l of OTS (octadecyltrichlorosilane) and of 60~ml
of toluene for
 about two hours. We then rinse the plane with chloroform. The
 AFM picture of surfaces prepared with this procedure is very
 similar to the one of plain pyrex planes (see Fig.\ref{AFM}).
The contact angles of macroscopic drops of water and glycerol have
been
 measured on each type of solid surfaces and have been referred in
 Tab.
\ref{tab:1}.
 In each case the hysteresis of the contact angle is about $2^{o}$.

 \begin{table}
\caption{Advancing contact angle of water and glycerol on plain
Pyrex and silanized Pyrex surfaces.} \label{tab:1}
\begin{tabular}{lll}
\hline\noalign{\smallskip}
System & advancing contact angle  \\
\noalign{\smallskip}\hline\noalign{\smallskip}
Glycerol-Pyrex & $<5^{o}$ \\
Glycerol-silanized Pyrex & $95^{o}$ \\
Water-silanized Pyrex & $100^{o}$ \\
\noalign{\smallskip}\hline
\end{tabular}
\end{table}

  After having mounted the surfaces on the apparatus, we put a drop of
  liquid between the sphere and the plane. In the experiments with
hydrophobic surfaces, only the plane is silanized. The
 experiments are carried out at ambient temperature, \textit{i.e.}, $25^{o}$.
The glycerol experiments are carried out in a dry atmosphere
 (some $\textrm{P}_{2}\textrm{O}_{5}$ is in the
same box than the DSFA).

\section{Hydrodynamic force induced by a drainage flow between a sphere and
a plane} \label{sec:dynameas}

\begin{figure}
\begin{center}
\includegraphics*{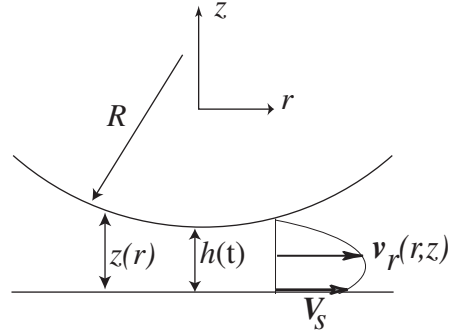} \caption{Drainage flow between a
sphere and a plane, in the case of a no-slip boundary condition on
the sphere and a partial slip of velocity $V_s$ on the plane.}
\label{figreynolds}
\end{center}
\end{figure}

 We consider here the flow of a liquid between a sphere with a radius $R$
and a plane. The
 sphere is moving in the direction normal to the plane as shown in
 Fig.\ref{figreynolds}.

 In the case of a no-slip boundary condition, the hydrodynamic force
$F_{v}$  between the sphere and the plane due to the viscous
dissipation is the so-called Reynolds force $F_v$, which acts in
the direction normal to the plane and reads:
\begin{equation}\label{Freyn}
F_{v}=\frac{6\pi\eta R^{2}}{h}\frac{\textrm{d}h}{\textrm{d}t}
\end{equation}
 where $\eta$ is the fluid viscosity, $h$ the distance
between the apex of the sphere and the plane and
$\textrm{d}h/\textrm{d}t$
 the sphere velocity. With the no-slip boundary condition, the
assumptions to obtain this result are that the liquid is
incompressible, the solid surfaces ideally rigid, the Reynolds
number of the flow is small, and the distance $h$ between the apex
of the sphere and the plane is small in comparison with $R$ so
that the lubrication approximation is valid.

In our experiments, we measure $h$ and $F$.
  As mentioned in the previous paragraph, the displacement of the
sphere towards the plane
 consists in two contributions: a quasi-static component $h_{dc}(t)$
and a sinusoidal component
$\widetilde{h}_{{ac}}e^{\textrm{\small{j}}\omega t}$.
$|\widetilde{h}_{{ac}}|$ is
  at most 5~\AA. Given this value,
   one can consider that the distance $h$ in Eq.\ref{Freyn}
reduces to $h_{dc}$. Therefore in what follows we do not make any
difference between these two notations. If the only force between
the surfaces is the viscous one due to the Reynolds force, we
have:
 $$\widetilde{G}(\omega)=\frac{\textrm{j}\omega 6\pi\eta
R^{2}}{h}$$
 $$\frac{1}{G^{\prime\prime}(\omega)}=\frac{h}{6\pi\eta
R^{2}\omega}$$ Thus in the case of a no-slip boundary condition,
the inverse of the damping $1/G^{\prime\prime}(\omega)$ varies
linearly with the distance $h$ between the surfaces, with a slope
$1/6\pi\eta\omega R^{2}$ which is related to the shear viscosity
$\eta$ of the liquid. The plot of $1/G^{\prime\prime}(\omega)$
versus $h$ is a straight line which intersects the $h$-axis at its
origin. If the elastic deformations of the surfaces are no longer
negligible, or if the fluid is not incompressible,
 a real part should be present in the dynamic signals. As long as the
dynamic signals remain purely imaginary, we can safely assume
 that those approximations hold.

If slippage occurs at the liquid-solid interface, the value
$V_{s}$ or the tangential flow velocity at the solid surface is
the slip velocity. If one assumes that the slip velocity is
proportionnal to the hydrodynamic stress acting at the wall, $V_s$
can be expressed as a function of the local shear rate in terms of
a slip length (or Navier length) $b$ \cite{Brillouin} :
\begin{equation}
\label{vgliss}
 V_{s}=b\frac{\textrm{d}v}{\textrm{d}z}
 \end{equation}
With a slip boundary condition, Eq.\ref{Freyn} is no longer
correct.
  Vinogradova \cite{Vino95} has calculated
 the expression of the viscous force between a sphere and a plane
with a partial slip HBC described by Eq.\ref{vgliss}  :
\begin{equation}
 \label{slipreynolds} \widetilde{F}_{ac}=\frac{\textrm{j} \omega
6\pi\eta R^{2}}{h}\widetilde{h}_{ac}f^{*}
 \end{equation}
 where,
$f^{*}=\frac{1}{4}[1+6\frac{h}{4b}((1+\frac{h}{4b})
\ln(1+\frac{4b}{h})-1)]$

In this case,  the plot of the inverse of the damping
$1/G^{\prime\prime}(\omega)$ versus $h$ is no longer a straight
line. However in the limit
 $h\gg b$, the following expression for the
asymptote is obtained ~:
\begin{equation}
\label{asympt}
 G^{\prime\prime}(\omega)=\frac{\omega 6\pi\eta
R^{2}}{h+b}
\end{equation}
Hence, for large distances, and in the case of partial slip at the
solid wall, the plot of  $1/G^{\prime\prime}(\omega)$ as a
function of $h$ is a straight line translated by a distance $b$
compared to the no-slip case and intersecting the \textit{h}-axis
at the value $h=-b$. This is equivalent to applying the no-slip
HBC inside the bulk of the solid at a distance $b$ of the surface,
which gives a physical meaning of the slip length.

It is of interest to calculate the shear rate to which the liquid
is submitted as a function of the sphere velocity and the gap $h$.
The actual shear rate in the fluid depends on the HBC.
Nevertheless, it is possible to estimate an effective shear rate
from the average velocity of the flow divided by the distance
between the surfaces. This effective shear rate is small on the
sphere-plane axis, and increases with the radial distance from the
axis up to a
 maximum value :
\begin{equation} \label{shearrate}
\dot \gamma_\textrm{\small{eff}} \equiv \left|\frac{\partial
v_r}{\partial r}\right|_\textrm{\small{max}}= \frac{9}{4}
\frac{(3R)^{1/2}}{(2h)^{3/2}}.\frac{\textrm{d}h}{\textrm{d}t}
\end{equation}
 obtained at the distance $r=(2Rh/3)^{1/2}$.
At larger distances from the axis, the effective shear rate decays
to zero. With $R=1$ mm, $f=30$ Hz,
$\textrm{Re}(\widetilde{h}_{ac})=1$~\r{A} we have, for $h=100$~nm,
$\dot \gamma_{\textrm{eff}}\simeq 30$~s$^{-1}$. In the following,
we use Eq.\ref{shearrate} to compare the typical shear rate
obtained in different experiments involving a squeezing flow
between a sphere and a plane  (SFA and modified AFM).

\section{Experimental results}
\label{sec:results}
\subsection{Wetting case : glycerol over Pyrex}
\begin{figure}
\begin{center}
\includegraphics*[height=6cm]{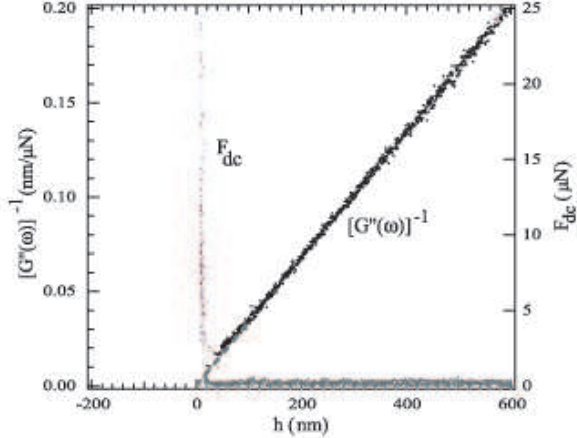}
\caption{Glycerol confined between plain Pyrex surfaces : static
force (grey) and inverse of the damping
$1/G^{\prime\prime}(\omega)$ (black) \textit{versus}
 the sphere-plane distance $h$. The dashed line is the best linear fit of $1/G^{\prime\prime}(\omega)$.
 The radius of the sphere is 1.5~mm and the frequency of the experiment is 39 Hz.} \label{glycm}
\end{center}
\end{figure}
 In Fig.\ref{glycm}, we have plotted on the same
 graph the evolution of the inverse of the damping
$1/G^{\prime\prime}(\omega)$,
 and of the static interaction force as a function of $h$, measured
for a glycerol film confined between Pyrex surfaces. The  measured
dynamic force $\widetilde{F}_{ac}$ is here out of phase with the
displacement excitation, so that a purely viscous force is
obtained. As predicted by the theory in the case of a no-slip HBC,
the plot of $1/G^{\prime\prime}(\omega)$ as a function of
 $h$ is a straight line. From the slope of this straight line we obtain
 the experimental value  $\eta$ = 350 mPa.s of the liquid viscosity,
 which corresponds to the viscosity of  glycerol containing a few percents of
water at $25^{o}$C. We have checked that this value is also
obtained when the viscosity of glycerol is measured with a
commercial rheometer under the same environmental conditions. The
intersection of this straigth line with the $h$-axis in
Fig.\ref{glycm} corresponds to the zero given by the mechanical
contact, \textit{i.e.} when the static interaction force raises
sharply. Therefore, we do not observe slip effects within the
range of a few nanometers in this system.

\subsection{Non wetting case}
\begin{figure}
\begin{center}
\includegraphics*[height=6cm]{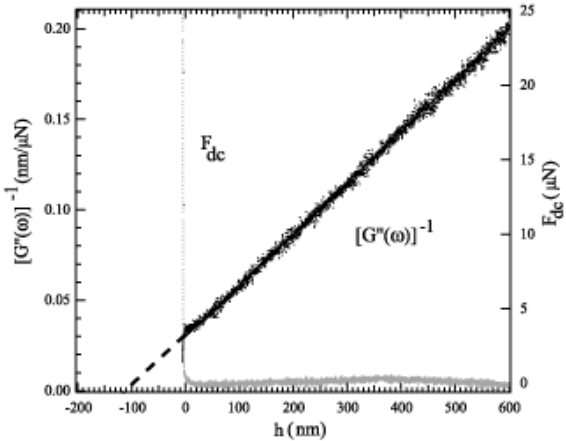}
\caption{Glycerol confined between a plain Pyrex sphere and a
silanized Pyrex plane: static force (grey) and inverse of the
damping $1/G^{\prime\prime}(\omega)$ (black) \textit{versus}
 the sphere-plane distance $h$. The dashed line is the best linear fit of $1/G^{\prime\prime}(\omega)$.
 The radius of the sphere is 1.5~mm and the frequency of the experiment is 39 Hz.} \label{glycnm}
\end{center}
\end{figure}
\begin{figure}
\begin{center}
\includegraphics*[height=6cm]{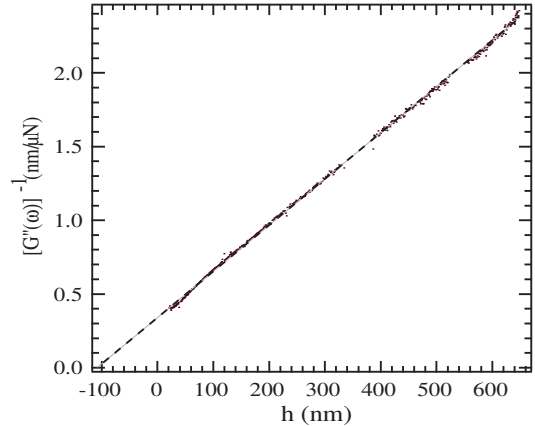}
\caption{Inverse of the damping $1/G^{\prime\prime}(\omega)$
induced by water confined between a plain Pyrex sphere and a
silanized Pyrex plane, \textit{versus} the sphere-plane distance
$h$.The dashed line is the best linear fit of
$1/G^{\prime\prime}(\omega)$. The radius of the sphere is 5~mm and
the frequency of the experiment is 79.5~Hz.} \label{eausil}
\end{center}
\end{figure}

In Fig.\ref{glycnm}, we have plotted the same quantities as in
Fig.\ref{glycm}, but in the case of a glycerol film flowing
between a silanized plane and a pyrex sphere. This is a typical
result of what is obtained in this non wetting system. The plot of
the inverse of the damping  as a function of $h$ is a straigth
line, whose slope is the same as that obtained in the wetting
case. This shows that there is no bulk effect associated to the
confinement of glycerol in this system. However the extrapolation
of the straight line intersects the $h$-axis at a negative
distance of about 100 nm from the contact. This decrease of the
hydrodynamic force shows the existence of liquid slippage occuring
at the hydrophobic solid surface. The order of magnitude of the
slip length is one hundred nanometers. Very similar hydrodynamic
forces are obtained in the case of water
 confined between the same kind of surfaces (see Fig.\ref{eausil}).
In all these experiments, the frequency response is independent of
the amplitude of excitation,  of the speed of the quasi static
motion of the surfaces (from 1 to 5 \AA/s) and proportional to the
excitation frequency (between 10 and 90 Hz). The hydrodynamic
force measured is the same wether the two surfaces are separated
from one another or brought together.

It must be stressed that we observe a significant variability of
the slip lengths measured on different hydrophobic samples
prepared with the same experimental procedure. The slip lengths
measured may vary between 50 to 200 nm, which is one order of
magnitude larger than the resolution of the experimental setup.
This variability occurs only with the non-wetting samples, and has
never been observed with Pyrex surfaces (wetting case).

Table \ref{tab_biblio} summarizes some experimental results
obtained by different research groups on the hydrodynamic boundary
condition of simple liquids on various solid surfaces. These
results show slip effects of different order of magnitude, with
slip lengths significanly lower
(\cite{Churaev84,Baudry2001,Craig2001}) or significatly larger
(\cite{Granick2001,tretheway2002}) than the ones obtained in this
study. The slip effects and slip lengths obtained in our
experiment are of the same magnitude as the one observed by Pit
\textit{et al.} \cite{Pit2000} with hexadecane over a bare
(wetting) or coated (partially wetting) sapphire surface. The
major difference lays in the influence of the wetting properties
of the surfaces. In our experiments slip occurs at the solid
surface only on non-wetting samples, whereas Pit \textit{et al}
observe significant slip of hexadecane on a wetting atomically
smooth sapphire surface. They also observe a no-slip condition on
a rough heterogeneous lyophobic surface. The surface roughness is
clearly an important parameter balancing the influence of wetting
properties on the HBC. Hydrodynamic calculations have shown that
for surfaces having locally the same HBC, the effect of roughness
is to decrease the effective slip undergone by the flow
\cite{Richardson73}. This effect of roughness may explain the
differences between Pit's results and ours regarding the influence
of wetting  properties on the HBC, although the slip lengths found
are of same magnitude. However SFA experiments of Chan and Horn
\cite{Horn85} on atomically smooth mica surfaces conclude on the
no-slip boundary condition. Another important parameter may be the
nature of the liquid. The results of Cheng \textit{et al} obtained
on surfaces whose wetting properties are unfortunately not
specified, show slippage effects which increase with the liquid
molecular size. The slip lengths found are of nanometer size,
significantly lower than ours and Pit's ones.

\begin{figure}
\begin{center}
\includegraphics*[height=6cm]{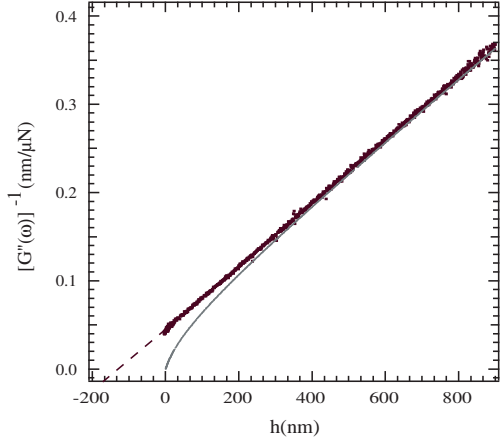}
\caption{Inverse of the damping $1/G^{\prime\prime}(\omega)$
\textit{versus} the sphere-plane distance $h$ : experimental
result obtained with glycerol in the non-wetting system (black),
and theoretical expression (Eq.\ref{slipreynolds}) corresponding
to a slip length b=180nm (grey). The dashed line is the best
linear fit of $1/G^{\prime\prime}(\omega)$ corresponding to the
asymptotic Eq.\ref{asympt}. The radius of the sphere is 1.3~mm and
the frequency of the experiment is 39~Hz.} \label{vino}
\end{center}
\end{figure}

A more detailed comparison of our experimental data with
macroscopic hydrodynamics shows that the hydrodynamic force
measured in the non-wetting system is in general not fully
compatible with a HBC corresponding to a well defined slip length
as described by Eq.\ref{vgliss}. Indeed the theoretical curve in
this case corresponds to the experimental data only in the
asymptotic limit $h\gg b$ (see Fig.\ref{vino}).  We do have no
full explanation for that discrepancy. Such discrepancies at small
distance have also been observed by Zhu and Granick
\cite{Granick2001} in SFA experiments using hydrophobic OTE coated
mica surfaces. They interpretate those discrepancies in terms of
an effective slip factor $f^*$ (see Eq.\ref{slipreynolds})
depending on the liquid flow rate. This corresponds to a
non-linear HBC. Such an interpretation is not possible in our
experiments, since the hydrodynamic force varies linearly with the
flow velocity, which can be varied either with the amplitude of
the excitation (1 to 5 \AA) or its frequency (10 to 90 Hz).
As far as the linearity of the HBC is concerned, it must be
emphasized that other experiments involving a squeezing flow
between a sphere and a plane with a partially wetting liquid-solid
system, performed by Craig \textit{et al.} \cite{Craig2001}, also
show non linear slip effects which appear only at high velocities
or small gap between the surfaces. It is however difficult to
compare the hydrodynamic parameters in the different experiments
with the sphere-plane geometry since they are done at different
velocities (velocities in \cite{Granick2001} vary between 4 and
400 nm/s,
 in \cite{Craig2001} between 1 to 20 $\mu$m/s, and in this work between
10 to 100 nm/s) but with different sphere radii (2 cm in
\cite{Granick2001}, 10 $\mu$m in \cite{Craig2001}, 1 mm in this
work). In order to compare these different flow situations, we
have indicated in Tab.\ref{tab_biblio} the range of the typical
shear rate in each experiment calculated from Eq. \ref{shearrate}.
It must be noted however that there is no experimental evidence
that this effective shear rate is the hydrodynamic parameter
controlling the occurrence of non-linear effects in the boundary
condition. Zu's experiments rather indicate a shear rate built on
the surfaces velocity divided by the gap between them.
Nevertheless the major difference between our experiment and the
ones of Craig et \textit{al.} and Zhu and Granick, is that we
clearly observe slip effects at large distances and independantly
of the flow velocity.

\onecolumn
\begin{table}
   \centering
{\tiny{
\begin{tabular}{l||c|c|c|c|c|c|c}
    \hline
Authors &Experimental& Solid surface & Liquids &Contact angle&
Typical
   &Results&Typical \\
   & method &  &  &or wettability&shear rate
   & & slip length \\
    &&  &  &($mJ/m^2$)&$\dot \gamma \ (s^{-1})$
   & &  \\
\hline \hline
   Churaev&Pressure& Quartz +&
   Hg/plain quartz& $115^o$ to $135^o$& 1-4
&Slip, linear HBC& 70$\pm$10~nm \\
    \cite{Churaev84}& drop&SAM of&Water   &
$70^o$ to $90^o$&  &Slip, linear HBC& 30$\pm$10~nm \\
     &  & trimethyl-& CCl$_4$  & $0^o$ & & No slip &
0$\pm$10~nm \\
        &  &-chlorosilane&Benzene  & $0^o$ & & No slip &
0$\pm$10~nm \\
   \hline
Cheng&Pressure& Glass \&
   & Water &not measured& 700-5000 &No-slip& 0~nm \\
      \cite{cheng2002}&  drop&photoresist & Hexadecane & "& &Slip,
linear HBC& 25~nm \\
                                        & &   & Decane&"
&  &Slip, linear HBC& 15~nm \\
                                        & & & Hexane&"
&  &Slip, linear HBC& 8~nm \\
    \hline
    Tretheway& PIV & Glass
& Water & $0^{o}$ &200& No-slip & 0~nm \\
    \cite{tretheway2002}& &+ SAM of OTS& " & $\theta\sim 120^{o}$ &&
Slip & 1~$\mu$m  \\
    \hline
Pit& FRAP& Bare saphire & hexadecane & $0^o$, $\gamma_{S}>72$&
10$^2$-10$^4$ &Slip, linear HBC& $175\pm 50$~nm \\
     \cite {Pit2000}& &+ SAM OTS & "&$0^o$, $\gamma_{S}=21$
& &Slip, linear HBC& $400\pm 50$~nm \\
   & &+ SAM FDS & "&$0^o$, $\gamma_{S}<13$
& & No slip & $0\pm 50$~nm \\
    \hline
Zhu &Dynamic& Bare mica & Tetradecane & $\theta=0^o$  &
50-15.10$^3$ & No-slip &
0~nm \\
                           \cite{Granick2001} & SFA&+ ads.surfactant&
Tetradecane  & $\theta=12^o$  &  & Non linear HBC&
0 to 1~$\mu$m \\
                                        & &+ SAM of OTE&
Tetradecane  & $\theta_a=44^o$   &   & Slip increases &
0 to 1.5~$\mu$m\\
                                          & &+ SAM of OTE&
Water & $\theta_a=110^o$  &   & with $\dot \gamma$ & 0 to 2.4~$\mu$m \\
\hline Chan &SFA& Bare mica &Hexadecane& $\theta=0^o$  & 50-500&
No-slip &
0~nm \\
                           \cite{Horn85} & &"&
Tetradecane& $"$  &  &No-slip &
"\\
                                     \hline
Baudry &Dynamic& Bare gold & Glycerol
& $\theta_a=60^o$ & 0.5-500& No-slip & 0~nm \\
    \cite{Baudry2001}&   SFA&+ SAM of thiols& " & $\theta_a=94^o$
&   &Slip, linear HBC& $\sim 40$~nm \\
    \hline
Craig &Modified& Gold & Sucrose solutions& $\theta_a=70^o$, &
30-10$^4$& Non linear HBC &
   0 to $\sim 15$~nm  \\
    \cite{Craig2001} & AFM&+ SAM of thiols&  & $\theta_r=40{^o}$ &   &Slip
increases&
    \\
&  &  &  &  &   &with $\dot{\gamma}$ & \\
    \hline
This &Dynamic&Bare pyrex & glycerol & $\theta=0^{o}$ &
0.5-500&  No-slip&  0~nm  \\
paper&SFA&  + SAM of OTS & Glycerol  & $\theta_a=95^{o}$ &
&Slip, linear
HBC&  50 to 200~nm  \\
   &  & + SAM of OTS & Water & $\theta_a=100^{o}$ &    &Slip, linear
HBC&  "  \\

\hline
\end{tabular}
}}\\
\caption{Summary of some slip lengths measurements obtained on
wetting and non-wetting surface by different research groups. The
experimental techniques involve pressure drop in flow through
capillaries, particle image velocimetry (PIV), fluorescence
recovery after photobleaching (FRAP), SFA and dynamic SFA, and AFM
with a microbead glued on the
 the cantilever tip. Lyophobic surfaces involve self-assembled
 monolayers (SAM) of octadecyltriethoxysiloxane (OTE),
 octadecyltrichlorosilane (OTS), perfluorodecanetrichlorosilane (FDS),
 and alcane-thiols. When specified in the references, the advancing
 and receding contact angles are noted $\theta_{a}$ and $\theta_{r}$.
 In SFA and modified AFM experiments, the maximum
 shear rates are calculated from Eq.\ref{shearrate}.
}\label{tab_biblio}

\end{table}

\twocolumn

\section{Conclusion}

We have investigated with a dynamic surface force apparatus the
flows of glycerol and water at a submicronic scale close to
hydrophobic surfaces. These experiments show the existence of slip
effects at the non-wetting solid wall which do not appear in flows
close to a wetting Pyrex surface. The scale of the slip effect
that we observe is of one hundred nanometers, and the hydrodynamic
force is fully linear with the liquid velocity. These results are
comparable qualitatively and quantitatively with other studies
reported in the literature with different or similar experimental
techniques \cite{Churaev84,Pit2000,Baudry2001}. They confirm that
the wetting character of the liquid-solid system is very important
for the flow properties at the solid interface.

However the hydrodynamic force is not fully compatible with an
hydrodynamic boundary condition corresponding to a well defined
slip length. We attribute this behavior to spatial heterogeneities
in the local hydrodynamic boundary conditions, over the scale
involved in the measurement of the hydrodynamic force with the
flow geometry of the experiment - typically 50~$\mu$m$\times
50$~$\mu$m. These spatial inhomogenities
 are specific to the  hydrophobic surfaces, but we have
not yet been able to correlate them to any direct observation with
AFM measurements. However, using taping mode atomic force
microscopy in water, Tyrell and Attard \cite{tyrell2001} have
shown that in some conditions hydrophobic surfaces in water may be
covered with flat nanobubbles extending laterally over hundreds of
nanometers. The presence of such flat nanobubbles or thin films of
air on hydrophobic surfaces, could be responsible for large slip
effects with significant spatial variability \cite{Stone2002}.
Further work is in progress for a better understanding of this
effect.

\vskip 1cm

We thank F. Feuillebois for helpful discussions, J. Klein for
enlightening discussions and communications, J.-P.~Rieu for help
in imaging the surfaces. We thank the DGA for its financial
support.

%
%
%

%
%
%

\end{document}